\documentclass[preprint,showpacs,nofootinbib,prc,aps]{revtex4-1}

\usepackage{graphicx,amstext,amssymb}

\begin{document}

\title{ Entanglement of scales  as a possible mechanism for  decoherence and
thermalization in relativistic heavy ion collisions }

\author{S.V. Akkelin$^{a}$, Yu.M. Sinyukov$^{a,b}$}

\affiliation{$(a)$ Bogolyubov Institute for Theoretical Physics, Metrolohichna str. 14b, 03680 Kiev, Ukraine \\ $(b)$ ExtreMe Matter Institute EMMI, GSI~Helmholtz~Zentrum f\"ur~Schwerionenforschung,
D-64291 Darmstadt, Germany}

\begin{abstract}
Despite the fact that a system created in relativistic heavy ion
collisions is an isolated quantum system, which cannot increase its
entropy in the course of  unitary quantum evolution, hydrodynamical
analysis of experimental data  seems to indicate that the matter
formed in the collisions is thermalized very quickly. Based on
common consideration of hydrodynamics as an effective theory in the
domain of slow- and long-length modes, we discuss the physical
mechanisms responsible for the decoherence and emergence of the
hydrodynamic behavior  in such collisions, and  demonstrate how such
physical mechanisms  work in the case of the scalar field model. We
obtain the evolution equation for the  Wigner function of a
long-wavelength subsystem  that describes its decoherence,
isotropization, and approach to thermal equilibrium induced by
interaction with short-wavelength modes.  Our analysis supports the
idea that decoherence, quantum-to-classical transition and
thermalization in isolated quantum systems are attributed to the
experimental context, and are related to a particular procedure of
decomposition of the whole quantum system into relevant and
irrelevant from an observational viewpoint subsystems.

\end{abstract}


\maketitle

\section{Introduction}

The unique and very ambitious program on the creation and study of a
small part of the early universe in laboratories   is being carried
out at the Relativistic Heavy Ion Collider (RHIC) and the Large
Hadron Collider (LHC). In collisions between  nuclei at these
machines a huge number of created particles forms a rapidly
expanding quark-gluon and/or hadron systems within space-time scales
$10^{-14}$ m and $10^{-22}$ s. One of the most important results
obtained  in these experiments is that hydrodynamic models with
nearly perfect fluid describe well the observables in relativistic
heavy ion collisions.   The best agreement with data is achieved  if
very early thermalization times ( $\lesssim 1 $ fm) of the produced
quark-gluon matter are assumed \cite{hydro-rev}.

In spite of a significant recent progress in the study  of early
time dynamics  of quark-gluon matter  produced in high energy heavy
ion collisions, thermalization  of the  matter created in $A+A$
collisions remains the great and still unresolved mystery (see,
e.g., Ref. \cite{Mull}). It seems that, at least partially, this  is
so because a system created in each  nucleus-nucleus collision can
be considered as an isolated quantum system that does not interact
with external environment, and unitary quantum evolution of an
isolated system cannot increase its entropy, no matter what happens
during this evolution (e.g., deconfinement, etc.), and no matter how
large and complex the system is.

To  avoid the conceptual problems with  thermalization of an
isolated quantum system, some approaches  utilize classical
approximation for early time dynamics in heavy ion collisions. We
mention here the approaches that are based on classical picture of
on-mass-shell particles: Boltzmann gas of particles with short-range
interactions (see, e.g., Ref. \cite{gas}), and plasma particles with
long-range interactions (see, e.g., Ref. \cite{Mrow}). The more
sophisticated approach for a description of early time dynamics and
equilibration processes in $A+A$ collisions utilizes the initial
state, which follows from the color glass condensate (CGC) effective
field theory (for recent reviews see, e.g., Ref. \cite{CGC} and
references therein). This  describes degrees of freedom in the
colliding nuclei as highly occupied gluon fields  with small gauge
coupling that are produced by the statistical ensemble of classical
color sources on an event-by-event basis. Such an initial state
corresponds to the Glauber coherent state \cite{Klauder} that
minimizes the uncertainty relation. Then, because for such a state
the classical $\hbar \rightarrow 0$ limit (see, e.g., Ref.
\cite{limit}) is equivalent to the  limit when the  coupling
constant tends to zero and the field's momentum and coordinate
expectation values tend to infinity, one  utilizes classical
Yang-Mills equations with fluctuating initial conditions as suitable
approximation  for the description of early time dynamics in
relativistic heavy ion collisions (see, e.g., Ref. \cite{Gelis-1}
and references therein).

It is still unclear whether or not such an approach can result in
proper early thermalization in $A+A$ collisions. Moreover, even if
the approach, based on the classical picture,  will result in early
thermalization, it does not help to understand this phenomenon in
$A+A$ collisions from the first principles. This  is so because a
 system created in an $A+A$ collision is inherently
quantum\footnote{For any fixed $\hbar$,  even if the classical and
quantum expectation values  coincide at the moment, they will
diverge from each other after  some time,  except for the specific
case of the Gaussian interaction (see, e.g., Ref. \cite{Got}).}, and
its initial state is, in fact, quantum superposition of the Glauber
coherent states. Only if the different  Glauber coherent states
could be distinguished experimentally as separate initial states  of
colliding nuclei, the initial condition  can be substituted by the
corresponding statistical mixture. If this  is not the case, then
such a substitution is unjustified. One more shortcoming with such
an approach is that unlike a classical system, where chaotic
behavior can appear due to extreme sensitivity of a complex system
to the initial conditions, in a quantum system the unitarity  of the
Schr\"odinger evolution preserves all scalar products and, so, all
the "distances" between quantum state vectors during the time, and
no chaotic behavior is possible. Therefore, even for a
quasiclassical initial condition the quasiclassical approximation
can be destroyed relatively quickly for systems that  exhibit
classical dynamical (deterministic) chaos. Indeed, it was
demonstrated that this happens after a time that  is only
logarithmic in the Plank constant \cite{Zurek-1} resulting in a
noticeable deviation between classical and quantum expectation
values \cite{Zurek-2}. Then, loosely speaking, the chaos seen in the
approximate classical dynamics of isolated quantum system is an
artifact of the approximations. Only decoherence caused by the
environment can substantially reduce the discrepancy between quantum
and classical expectation values and restore the quantum-classical
correspondence for  a classically chaotic system
\cite{Zurek-1,Zurek-2,Zurek-3}.

Perhaps, a hope that the anti-de-Sitter/conformal-field-theory
(AdS/CFT) correspondence \cite{Ads} (for reviews see Refs.
\cite{Ads-rev1,Ads-rev2}) can help to understand thermalization in a
closed quantum system is one of the reasons why this approach, which
attempts to explain the origin of decoherence and thermalization in
$A+A$ collisions, has recently attracted much attention in the heavy
ion community. While AdS/CFT correspondence does not take place  for
QCD, it is generally believed that it provides correct qualitative
picture of QCD dynamics in the strongly coupled
regime.\footnote{Note here, that the  applicability of such a regime
for the early stage dynamics in relativistic heavy ion collisions is
questionable, see, e.g., Ref. \cite{Blaizot-1}.} The AdS/CFT
correspondence is based on the holographic gauge/string duality
between  four-dimensional (4D) quantum field gauge theory such as $N
= 4$ super Yang-Mills gauge theory (which is a conformal field
theory), and five-dimensional quantum string theory. Therefore, the
AdS/CFT correspondence is also sometimes called a gauge/string
duality. The duality means an exact equivalence between two
theories, i.e., it means that any calculated quantity can be
expressed in terms of a dual partner theory. In practice, however,
calculations in dual 5D quantum string theory are possible only
under some limitations, which  from the QCD viewpoint means that
$N_{c}\rightarrow \infty$ and $\lambda \rightarrow \infty$, where
$N_{c}$ is the number of colors and $\lambda$ is the QCD coupling
constant. Under such conditions a gauge/string duality is reduced to
a gauge/gravity duality between 4D quantum gauge theory and 5D
classical gravity theory. In this correspondence, the radial
coordinate $r$ of additional spatial direction can be associated
with the renormalization group energy scale (energy cutoff scale) in
the gauge field theory \cite{RG}, and asymptotically high values of
radius parameter correspond to gauge field theory with
asymptotically high energy cutoff. Therefore, the AdS/CFT
correspondence  can be treated as geometrization of a
renormalization group.

The phenomenon of thermalization of 4D quantum field theory in this
approach is then associated with the irreversible process of  black
hole (and corresponding event horizon with non-zero entropy)
formation \cite{Ads-rev2}. Specifically,  it was found that the
long-wavelength (smoothed over short-scales) approximation of
metrics induced by a large stationary black hole corresponds to a
thermal state of the gauge quantum field theory, and the
long-wavelength approximation of metrics induced by a large nearly
stationary black hole corresponds to a nearly perfect hydrodynamical
structure  of the expectation value of the energy momentum tensor of
gauge quantum fields. The latter duality is sometimes named as the
fluid/gravity correspondence \cite{Hubeny}. Such a fluid/gravity
correspondence is a useful tool to calculate viscosity  for strongly
interacting locally equilibrated systems \cite{Ads-rev2}. But the
question appears: does such duality explain the thermalization
process in $A+A$ collisions?

First, note that sometimes, to support idea of "black hole
thermalization", one appeals to the well known and well understood
Unruh effect \cite{Unruh-1} (for review see, e.g., Ref.
\cite{Unruh-rev}). It is noteworthy that the dual description of
this effect from the point of view of an accelerating observer and
inertial observer gives the same result: thermalization of an
uniformly accelerated (in inertial reference system) particle
detector \cite{Unruh-2,Unruh-rev} due to quantum interactions.
However, one needs to remember that conditions, for which the Unruh
effect takes place, mean that the accelerated detector is an open
system that is accelerated by some external forces, and such a
thermalization cannot be observed for a closed isolated system with
an  accelerated subsystem -- detector \cite{Sinyukov}. This point
was missed in recent attempts to explain thermalization in $A+A$
collisions by means of the Unruh mechanism \cite{Kharzeev}.

Second, note that AdS/CFT correspondence cannot be treated as the
origin of thermalization and entropy creation in dual 4D quantum
field theory  because the latter  is an ordinary quantum field
theory in flat space-time, and so cannot produce entropy in the
course of reversible and unitary quantum evolution. Then, based on
general principles of quantum theory one can infer that the
fluid/gravity correspondence  is valid for some  decomposition of
the whole quantum system into separate subsystems: it is well known
that while entropy of the whole isolated quantum system remains
constant under the time evolution, entanglement entropies of its
subsystems can increase. Indeed, recent studies in time-dependent
AdS/CFT based on a holographic formula of the entanglement entropy
\cite{Tak-1} demonstrate that black-hole formation in AdS dual can
be associated with an increase of the entanglement entropy in CFT
dual \cite{Tak-2} (for recent reviews see Ref. \cite{Tak-3}), but
the latter takes place only after splitting of  CFT dual into
spatially separated subsystems with  a quantum quench of one of them
at a specific instant of time.

In what follows, we adopt the standpoint that (entanglement) entropy
production in an isolated quantum system can take place only after
its decomposition into subsystems, and that the specific way of
separation of the closed system into subsystems depends on a certain
experimental context (i.e., it is related with "relevant"
observables). Instead of decomposition of the system into
separated-in-space subsystems, we split the system into
long-wavelength modes and short-wavelength modes  subsystems, and
treat the long-wavelength modes subsystem as the  relevant one and
the  short-wavelength modes subsystem as the environment. Such a
decomposition seems to be appropriate for the experimental context
in $A+A$ collisions, and is in agreement  with common consideration
of hydrodynamics as the effective theory in the domain of slow- and
long-length modes (see, e.g., Ref. \cite{ads-hyd}). Inasmuch as  the
aim of this paper is partly methodological, we will focus mainly on
the general quantitative features of the evolution of long
wavelength quantities in  a simple scalar field theory model to
investigate the physical mechanisms responsible for the decoherence
and emergence of the hydrodynamic behavior in $A+A$ collisions.

\section{Decoherence  and  approach to equilibrium of the long wavelength  observables}

Let us start with pointing out that  necessary condition for
emergence of hydrodynamic behavior is the  decoherence, i.e.,
suppression of interference of some  set of variables  that allows
one to use local densities to describe a system's dynamics.  It is
well known that an open system can be decohered (i.e., its state can
be approximately diagonalized in some basis) and can acquire
classical properties due to interactions with its environment
containing the many degrees of freedom that are ignored from an
observational point of view (for review see \cite{Zeh-1} and
references therein). Note that such a decoherence due to
interactions between the system and its environment is formulated
entirely quantum mechanically, and globally the quantum mechanical
superposition remains unchanged, as required by the unitarity of the
evolution of the total wave function. In contrast with such an
environment-induced decoherence, in a relativistic nucleus-nucleus
collision the system remains isolated after preparation and until an
observation at a large time $t_{out}$ is performed. However, it is
well known that while decoherence of an isolated system is
impossible, the decoherence of its subsystems is still possible:
while the state of the whole system remains pure, the state of a
subsystem of a composite system can be described as improper mixture
\cite{improper} represented by the partial trace of the statistical
operator of the composite system in a pure state (proper mixture
means incomplete knowledge for a pure state, and, typically,
represents a statistical ensemble). The key point here is quantum
entanglement: interacting quantum subsystems become entangled in the
course of unitary evolution of the system as a whole and, as a
result, the quantum states of subsystems become mixed states. It is
noteworthy that  such mixed states generation has nothing to do with
the  formation of statistical ensembles when the weights of the
states have  no relation to the exact dynamical equations.

Note that  because of the quantum non-separability \cite{improper} a
closed system can be resolved into parts ("subsystems") in various
ways.  Different splittings result in complementary descriptions of
a system, and the state of a whole quantum system can not be
inferred from the states of its parts unlike the state of a
composite classical system, which  can always be reconstructed from
the known  states of its parts. Decoherence and, perhaps,
thermalization thus arise from the description of the system by an
observer who at the selected measurements and data analysis has
access only to subsystem degrees of freedom, while residual degrees
of freedom  are entangled with the subsystem but remain unobserved.
The state of the whole system, however, remains pure, and its
entropy  remains zero: due to the quantum entanglement, the entropy
of a whole quantum system is not equal to the  sum of the entropies
of its parts that are defined as the von Neumann entropies of the
corresponding reduced density matrices. In this kind of process, the
equilibrium state of the relevant subsystem  is just a state when
its entropy reaches a maximum due to the build-up of entanglement of
the considered system with its environment induced by the
interactions \cite{Gemmer}.

Then, to explain the success of hydrodynamics in $A+A$ collisions,
one can assume that a system created in a relativistic $A+A$
collision can be decomposed into the fast short-length modes that
represent irrelevant (i.e., observationally inaccessible or ignored)
degrees of freedom, and slow long-length modes that represent
relevant (i.e., observationally accessible) degrees of freedom. The
former act as environment and can ensure decoherence and approach to
(local) equilibrium for the latter \cite{Calzetta} (see also Ref.
\cite{Elze-1}). Such a splitting is conditioned by the experimental
context because of limited region and accuracy in a measurement of
relevant observables (e.g., particle momentum spectra) and, also,
because not all possible observables are measured (e.g., not  all
$N$-particle correlations, quantum interference effects, etc.).

The evolution  of the relevant subsystem of closed system is studied
usually  by means of utilizing  powerful mathematical methods, e.g.,
by means of the projection operator technique (for review see, e.g.,
Ref. \cite{Rau-1}).  Note, however, that application of these
methods to non-equilibrium quantum field theory is usually rather
complicated and sometimes ambiguous, and physics is often hidden by
mathematical formalisms. Therefore, for illustrative purposes, we
will utilize here a more heuristic coarse graining approach  aiming
to make clear origin of decoherence of relevant observables in $A+A$
collisions and their subsequent evolution towards  equilibrium.

Due to the complexity of the problem, we restrict ourselves to a
$\varphi^4$ quantum field model, whose dynamics is determined by the
Lagrangian density
\begin{eqnarray}
L = \frac{1}{2}  \partial_{\mu} \varphi
\partial^{\mu} \varphi  - \frac{\lambda}{4 !} \varphi^4, \label{1}
\end{eqnarray}
where $\lambda$ is coupling constant.   In the following, we utilize
the Heisenberg representation. Expectation values are defined as
$\langle O \rangle = Sp (\hat{\rho} O)$, where $\hat{\rho}$ denotes
the statistical operator associated with an initial (pure) state  of
the system.

The expectation value of the energy momentum tensor, $\langle
T_{\mu\nu} \rangle$,   satisfies to conservation equations
\begin{eqnarray}
\partial^{\mu} \langle T_{\mu\nu} \rangle =0, \label{2}
\end{eqnarray}
which  follows from the field evolution equation. Many studies of
the $\langle T_{\mu\nu} \rangle$ evolution were based on classical
field approximation of the energy momentum tensor,
\begin{eqnarray}
\langle T^{\mu\nu}(x)\rangle  \approx
T^{\mu\nu}[\langle\varphi\rangle]=
\partial^\mu \langle\varphi\rangle \partial^\nu\langle\varphi\rangle
-
g^{\mu\nu}\,\Big[\frac{1}{2}(\partial_\alpha\langle\varphi\rangle)^2-\frac{\lambda}{4!}\langle\varphi\rangle^4\Big].
\label{3}
\end{eqnarray}
It is worth to note here that  such an approximation does not mean
that the evolution of $\langle T_{\mu\nu} \rangle$ proceeds as in
classical field theory. This  is so because  the expectation value
of the field, $\langle \varphi \rangle$, is governed by the equation
\begin{eqnarray}
\partial^\mu\partial_\mu\langle\varphi\rangle =
- \frac{\lambda}{3 !}\langle \varphi^3 \rangle, \label{4}
\end{eqnarray}
where  $\langle \varphi^3 \rangle $ in the right hand side contains
correlations of quantum fluctuations. Classical evolution for
$\langle T_{\mu\nu} \rangle$ can be obtained  if  $\langle \varphi^3
\rangle$ is approximated by  $\langle \varphi \rangle^3$, which
leads to the  reversible classical evolution equation for $\langle
\varphi \rangle$.  It is well known, however, that classical
equations approximate the underlying microscopic quantum dynamics
for the very special initial coherent (Glauber) state \cite{Klauder}
and during a limited time period only. On the other hand, if instead
of the whole system  we consider the  relevant subsystem, and
associate the latter with long wavelength modes (i.e., with momentum
scales $k$ smaller than the some characteristic scale $k^\star$),
then utilization of the classical approximation for expectation
values of long wavelength observables can be justified, and quantum
fluctuations can be accounted for short-length modes
only.\footnote{Note that quantum correlations are suppressed for the
long-wavelength modes because long-wavelength mode operators are, in
fact, smeared operators, and the canonical commutation relation for
the smeared conjugated operators tends to zero if the scale of
averaging tends to infinity (see, e.g. Ref. \cite{Vikman} and
references therein). This allows one to use classical approximation
the long-wavelength modes evolution after decoherence, the latter is
necessary but not sufficient condition for classical
approximations.}

Let us split the  quantum field $\varphi $ at $t=t_0$ into
long-wavelength modes $ \varphi _L^{t_{0}}$, and short-wavelength
modes $ \varphi _S^{t_{0}}$: $ \varphi $= $ \varphi _{L}^{t_{0}} +
\varphi _{S}^{t_{0}}$. We assume that the initial long wavelength
field, $\varphi_{L}^{t_{0}}$, corresponds to a convolution of field
operator  $\varphi $ with a "window" function $W_V$, $\int W_{V} =
1$, which makes smoothing/averaging   of the field over a domain of
size $V=1/k^{*3}$,
\begin{eqnarray}
\varphi_{L}^{t_{0}}  (x,t_{0})= \int d^3 x' W_{V}(x-x') \varphi
(x',t_{0}). \label{5}
\end{eqnarray}
Also, we split a state of the system into $L$ and $S$ subsystems:
$\hat{\rho}_{L}\otimes \hat{\rho}_{S}$,  assuming that observables
correspond to operators acting on $L$ states only.

The evolution equation for expectation value of long-length modes
with  initial condition defined according to (\ref{5})  reads
\begin{eqnarray}
\partial^\mu\partial_\mu\langle \varphi_{L}^{t_{0}} \rangle= - \partial^\mu\partial_\mu\langle \varphi_{S}^{t_{0}} \rangle
- \frac{\lambda}{3 !}\langle
(\varphi_{L}^{t_{0}}+\varphi_{S}^{t_{0}})^3 \rangle . \label{5.1}
\end{eqnarray}
One can see that in the  course of evolution the initially smeared
field becomes dependent on short-wavelength modes. This  is a
manifestation of quantum entanglement in the Heisenberg picture. To
follow the evolution of the corresponding observables, one needs to
make repeated in time splitting of the whole quantum system into the
corresponding subsystems, in the Heisenberg picture this  means that
one needs to make  repeated redefinition of the corresponding
observables (this is reminiscent of the familiar repeated randomness
assumption in the Boltzmann kinetics).

Then,  to calculate observables associated with long wavelength
modes, one needs to supplement this exact motion with an operation
that prevents the state to deviate too much from $L$. This  can be
done by dividing the evolution into time intervals, and choosing
initial conditions for each time step  with $\varphi_{L}^{t_{i}} $
being replaced at the time $t_{i+1}=t_{i}+\delta t$ by the
associated $ \varphi_{L}^{t_{i+1}}  (x,t_{i+1})=  \int d^3 x'
W_{V}(x-x') ( \varphi_{L}^{t_{i}} (x',t_{i+1})+ \varphi_{S}^{t_{i}}
(x',t_{i+1}))$. Then for $t_{i+1}<t<t_{i+2}= t_{i+1}+\delta t$,
\begin{eqnarray}
\partial^\mu\partial_\mu\langle \varphi_{L}^{t_{i+1}} \rangle= - \partial^\mu\partial_\mu\langle \varphi_{S}^{t_{i+1}} \rangle
- \frac{\lambda}{3 !}\langle
(\varphi_{L}^{t_{i+1}}+\varphi_{S}^{t_{i+1}})^3 \rangle ,\label{5.2}
\end{eqnarray}
and we have piecewise continuous description of $L$-modes
evolution.\footnote{Note that an exact equation of motion for the
relevant variables may be obtained by this procedure if their
characteristic time scale are much larger than the time scales
associated with the irrelevant variables and if the time $\delta t$
is chosen in between \cite{Balian-1}.} Now, let us neglect in each
$\delta t$-interval contribution of long-scale quantum fluctuations
and  contribution of the  short wavelength modes into the right-hand
side of the evolution equations. Then we get the chain of equations
\begin{eqnarray}
\partial^\mu\partial_\mu\langle \varphi_{L}^{t_{0}} \rangle =
- \frac{\lambda}{3 !}\langle \varphi_{L}^{t_{0}} \rangle^{3}, t_{0}<t<t_{1},  \label{5.3} \\
...\nonumber \\
\partial^\mu\partial_\mu\langle \varphi_{L}^{t_{n}} \rangle =
- \frac{\lambda}{3 !}\langle \varphi_{L}^{t_{n}} \rangle^{3},
t_{n-1}<t<t_{n}, \label{5.4}
\end{eqnarray}
which  approximates piecewise continuous description of $L$-modes
till some time $t_{n}$. Note  that  the projection times set,
$\{t_i\}$, is not uniquely defined and can  vary in some intervals
allowed by dynamics. Therefore, such a piecewise continuous
description means that we have, in fact, a set of different
histories of the $L$-modes evolution with randomly chosen projection
times and, so, random expectation values of $L$-modes. Such a  set
of piecewise continuous evolutions  can be approximated   by the
continuous one,
\begin{eqnarray}
\partial^\mu\partial_\mu \langle\varphi_{L}\rangle_{\xi}=
- \frac{\lambda}{3 !}(\langle \varphi_{L} \rangle_{\xi}^{3} + \xi),
\label{6}
\end{eqnarray}
where $\xi$ accounts for random discontinuity $\langle
\varphi_{L}^{t_{i}} \rangle(t_{i+1})\neq \langle
\varphi_{L}^{t_{i+1}} \rangle (t_{i+1})$ and, so, is associated with
 fluctuations of the expectation value of long wavelength modes.
As we discussed above, such a discontinuity is caused  by the
interaction of long wavelength modes with the short wavelength ones,
in particular, by the interaction with the short-scale quantum
fluctuations that typically are more enhanced than  the long-scale
ones. Because the information transferred towards the irrelevant
variables is discarded at the beginning of each time interval, $\xi$
becomes a stochastic "noise" variable, and induces a continuous time
random walk stochastic dynamics for
$\langle\varphi_{L}\rangle_{\xi}$. Then, to get true long wavelength
observables without "trembles" that are associated with different
projection histories, one needs to average such observables  over
$\xi$. Such an averaging means, in fact, smearing over the time
interval $\delta t$ for set of projection histories, and is not
associated with statistical ensemble of initial events. The
necessary condition for hydrodynamical approximation  to be valid is
the allowance to neglect, after such an averaging, non-conservation
of energy momentum due to interactions with short wavelength modes,
i.e., to make the  assumption that such an interaction results
mostly in the information loss.

Direct calculation of $ T_{\mu\nu}[\langle\varphi_{L}\rangle_{\xi}]
$ based on evolution equations for $\langle\varphi_{L}\rangle_{\xi}$
is a rather uneasy task, which can hardly be done analytically.
Therefore, here we proceed  in a more heuristic way and express $
T_{\mu\nu}[\langle\varphi_{L}\rangle_{\xi}] $ through the
expectation value of the corresponding Wigner operator (see, e.g.,
Ref. \cite{Groot}), and obtain for the latter a kinetic transport
equation. Let us define the (reduced) Wigner function describing the
state of the long-wavelength modes, $ N_{L}(x,p)$, as
\begin{eqnarray}
N_{L}(x,p)= \sum_{\xi} N_{L}^{\xi}(x,p), \label{8.0}
\end{eqnarray}
where
\begin{eqnarray}
N_{L}^{\xi}(x,p)=(2\pi)^{-4}\int d^{4}v
e^{-ipv}\langle\varphi_{L}\rangle_{\xi}(x+\frac{1}{2}v)\langle\varphi_{L}\rangle_{\xi}(x-\frac{1}{2}v),
\label{8}
\end{eqnarray}
and symbol $\sum_{\xi}$ means that we perform  in (\ref{8.0}) the
average with respect to random $\xi$ fluctuations, as was discussed
above.

Then the energy momentum tensor of long wavelength modes,
$T_{\mu\nu}^{L}(x)$, can be defined as averaged over $\xi$ classical
approximation of $\langle T_{\mu\nu}\rangle_{L}$:
\begin{eqnarray}
\langle T_{\mu\nu}\rangle_{L} \approx T_{\mu\nu}^{L}(x)= \sum_{\xi}
T_{\mu\nu}[\langle\varphi_{L}\rangle_{\xi}], \label{9.0}
\end{eqnarray}
where $T_{\mu\nu}[\langle\varphi_{L}\rangle_{\xi}]$ is written as in
(\ref{3}) but with substitution $\langle\varphi\rangle \rightarrow
\langle\varphi_{L}\rangle_{\xi}$. By means of the Wigner function
(\ref{8}) one can rewrite
$T_{\mu\nu}[\langle\varphi_{L}\rangle_{\xi}]$  as \cite{Groot}
\begin{eqnarray}
T_{\mu\nu}[\langle\varphi_{L}\rangle_{\xi}]
&=&\int\mbox{d}^4p\;\left (p_\mu
p_\nu+\frac{1}{4}\partial_{x^\mu}\partial_{x^\nu}
-\frac{1}{2}g_{\mu\nu}(p^2+\frac{1}{4}\partial_x^{\;2})\right )
N_{L}^{\xi}(x,p) \nonumber \\  &\;&+\frac{\lambda g_{\mu\nu}}{4!}
\int\mbox{d}^4p \mbox{d}^4p' N_{L}^{\xi}(x,p)N_{L}^{\xi}(x,p').
\label{9}
\end{eqnarray}
Using (\ref{6}), we obtain the following time evolution of the
Wigner function:
\begin{eqnarray}
p_{\mu}\partial^{\mu}N_{L}(x,p)= \nonumber
\\ \frac{i}{2(2\pi)^{4}}\sum_{\xi} \int d^{4}v
e^{-ipv}\left(\rho_{\xi}(x-\frac{v}{2})\langle\varphi_{L}\rangle_{\xi}
(x+\frac{v}{2}) -   \langle\varphi_{L}\rangle_{\xi}
(x-\frac{v}{2})\rho_{\xi} (x+\frac{v}{2}) \right ). \label{10}
\end{eqnarray}
Here
\begin{eqnarray}
\rho_{\xi}= - \frac{\lambda}{3 !}( \langle\varphi_{L}\rangle_{\xi}^3
+ \xi). \label{10.0}
\end{eqnarray}
Aiming to derive the kinetic equation for the Wigner function, let
us rewrite the above equation in the form
\begin{eqnarray}
p_{\mu}\partial^{\mu}N_{L}(x,p)= \nonumber
\\ \frac{i}{2(2\pi)^{4}}\sum_{\xi}
\int d^{4}v
e^{-ipv}\langle\varphi_{L}\rangle_{\xi}(x+\frac{v}{2})\langle\varphi_{L}\rangle_{\xi}
(x-\frac{v}{2}) \left(\varrho_{\xi}(x-\frac{v}{2})  - \varrho_{\xi}
(x+\frac{v}{2}) \right ), \label{10.1}
\end{eqnarray}
where
\begin{eqnarray}
\varrho_{\xi} = \frac{\rho_{\xi} }{
\langle\varphi_{L}\rangle_{\xi}}. \label{10.2}
\end{eqnarray}
Then, performing the Tailor expansion of
$(\varrho_{\xi}(x-\frac{v}{2}) - \varrho_{\xi} (x+\frac{v}{2})  )$
in powers of $v$  and integrating over $v$, we get
\begin{eqnarray}
p_{\mu}\partial^{\mu}N_{L}(x,p)=\frac{1}{4}\sum_{\xi}
\partial^{\mu}\varrho_{\xi} (x) \frac{\partial}{\partial p^{\mu}} N_{L}^{\xi}(x,p) + \sum_{\xi}\Phi_{\xi}(x,p), \label{11}
\end{eqnarray}
where we used (\ref{8}) and  $\sum_{\xi}\Phi_{\xi}(x,p)$ includes
all high derivatives terms of the Tailor expansion. Let us make the
natural assumption that averaging over $\xi$ reduces high
derivatives terms and allows one to neglect the last term in Eq.
(\ref{11}). Then in such an approximation
\begin{eqnarray}
p_{\mu}\partial^{\mu}N_{L}(x,p)=\frac{1}{4}\sum_{\xi}
\partial^{\mu}\varrho_{\xi} (x) \frac{\partial}{\partial p^{\mu}} N_{L}^{\xi}(x,p), \label{11.1.1}
\end{eqnarray}
and $N_{L}^{\xi}(x,p)$, as follows from (\ref{8.0}), is governed by
the equation
\begin{eqnarray}
p_{\mu}\partial^{\mu}N_{L}^{\xi}(x,p)=\frac{1}{4}
\partial^{\mu}\varrho_{\xi} (x) \frac{\partial}{\partial p^{\mu}} N_{L}^{\xi}(x,p).\label{11.0}
\end{eqnarray}
Let us define
\begin{eqnarray}
\delta N_{L}^{\xi} &= & N_{L}^{\xi} - N_{L}, \label{11.1} \\
\delta \varrho_{\xi} &=& \varrho_{\xi} - \varrho, \label{11.2}
\end{eqnarray}
here $\varrho= \sum_{\xi} \varrho_{\xi}$. Then, using (\ref{11.1})
and (\ref{11.2}),  Eq. (\ref{11}) reads
\begin{eqnarray}
p_{\mu}\partial^{\mu}N_{L}(x,p)=\frac{1}{4}
\partial^{\mu}\varrho (x) \frac{\partial}{\partial p^{\mu}} N_{L}(x,p)
+ \frac{1}{4} \sum_{\xi}[\partial^{\mu}\delta  \varrho_{\xi} (x)
\frac{\partial}{\partial p^{\mu}}\delta N_{L}^{\xi}(x,p)].
\label{11.3}
\end{eqnarray}
Now one needs to calculate the second term in the right-hand side of
the above equation. Subtracting (\ref{11.3}) from (\ref{11.0}) and
keeping only the lowest terms in $\delta$, one can get
\begin{eqnarray}
p_{\mu}\partial^{\mu}\delta N_{L}^{\xi}(x,p)=\frac{1}{4}
\partial^{\mu}\delta \varrho_{\xi} (x) \frac{\partial}{\partial p^{\mu}} N_{L}(x,p) + \frac{1}{4}
\partial^{\mu}\varrho (x) \frac{\partial}{\partial p^{\mu}} \delta N_{L}^{\xi}(x,p), \label{11.4}
\end{eqnarray}
which  can be rewritten as
\begin{eqnarray}
\delta N_{L}^{\xi} (x,p)=\delta N^{(free)\xi}_{L} (x,p)+
\frac{1}{4}\int d^{4}y G_{p}(x-y) \delta
\varrho^{\mu}_{\xi}(y)\frac{\partial}{\partial
p^{\mu}}N_{L}(y,p) + \nonumber \\
\frac{1}{4}\int d^{4}y G_{p}(x-y)
\varrho^{\mu}(y)\frac{\partial}{\partial p^{\mu}} \delta
N_{L}^{\xi}(y,p). \label{12}
\end{eqnarray}
Here $\varrho^{\mu}\equiv\partial^{\mu}\varrho$,
$\delta\varrho^{\mu}_{\xi}\equiv\partial^{\mu}\delta \varrho_{\xi}$,
 $\delta N^{(free)\xi}_{L} (x,p)$ is the general solution of the
homogeneous equation:
\begin{eqnarray}
p_{\mu}\partial^{\mu}\delta N^{(free)\xi}_{L} (x,p)(x,p)=0,
\label{12.1}
\end{eqnarray}
and
\begin{eqnarray}
p^{\mu}\partial_{\mu}G_{p}(x)=\delta^{(4)}(x) , \label{green1} \\
G_{p}(x)=p_{0}^{-1}\Theta (t)
\delta^{(3)}(\mathbf{r}-(\mathbf{p}/p_{0})t). \label{green2}
\end{eqnarray}
Let us assume that initially $\delta N^{\xi}_{L}=0$. Then $\delta
N^{(free)\xi}_{L}=0$ and one can see from Eq. (\ref{12}) that in
lowest order in $\delta$
\begin{eqnarray}
\delta N_{L}^{\xi} (x,p)= \frac{1}{4}\int d^{4}y G_{p}(x-y) \delta
\varrho^{\mu}_{\xi}(y)\frac{\partial}{\partial p^{\mu}}N_{L}(y,p).
\label{13}
\end{eqnarray}
In such an approximation  (\ref{11.3}) reads
\begin{eqnarray}
p_{\mu}\partial^{\mu}N_{L}(x,p)= \nonumber \\ \frac{1}{4}
\partial^{\mu}\varrho(x) \frac{\partial}{\partial p^{\mu}} N_{L} (x,p)+  \frac{1}{16}
\frac{\partial}{\partial p^{\mu}} \int d^{4}y
G_{p}(x-y)\sum_{\xi}[\delta \varrho^{\mu}_{\xi}(x)\delta
\varrho^{\nu}_{\xi}(y)] \frac{\partial}{\partial p^{\nu}}N_{L}(y,p).
\label{14}
\end{eqnarray}
In general, we cannot compute exactly the contributions of the
fluctuations (otherwise we could solve exactly the model):
approximations are necessary. Then, to proceed further we have to
specify the stochastic properties of the random quantities $\delta
\varrho^{\mu}_{\xi}(x)$. We take the simplest ansatz assuming that
the backreaction is negligible
\begin{eqnarray}
\sum_{\xi}[\delta \varrho^{\mu}_{\xi}(x)\delta
\varrho^{\nu}_{\xi}(y)] = \tau^{\mu \nu}(x,y)  \delta (t_{x}-t_{y}).
\label{15}
\end{eqnarray}
The assumption of a $\delta-$function in time difference
 means that the auto-correlation time of the
fluctuations is small compared to the time scale of the  motion of
the averaged fields. The fluctuations thus appear as uncorrelated on
the time scale of the motion of the averaged fields. This assumes a
clear separation of  scales between the short time scale of
irrelevant degrees of freedom, and the long  time scale  which
characterizes the dynamics of the relevant degrees of freedom.

Then
\begin{eqnarray}
p_{\mu}\partial^{\mu}N_{L}(x,p)=\frac{1}{4}
\partial^{\mu}\varrho (x) \frac{\partial}{\partial p^{\mu}} N_{L} (x,p)+  \frac{1}{16}\tau^{\mu \nu}
\frac{\partial}{\partial p^{\mu}} \frac{1}{p_{0}}
\frac{\partial}{\partial p^{\nu}}N_{L}(x,p), \label{16}
\end{eqnarray}
here $\tau^{\mu \nu}\equiv \tau^{\mu \nu} (x,x)$. As usual, the
irreversible transport equation for relevant subsystem is valid only
for finite time scales where the short-memory approximation (i.e.,
"white noise" approximation) is justified.  It is worth to note the
similarity  of  Eq. (\ref{16}) with the Fokker-Plank equation, the
latter is often utilized for a description of  the approach to
(local) equilibrium. In the utilized approximation, see Eqs.
(\ref{5.4}) and (\ref{6}), we do not account for explicit
contribution of short-wavelength modes, so $\sum_{\xi}\xi =0 $ and
$\varrho= \sum_{\xi} \varrho_{\xi}=- \frac{\lambda}{3 !}
\langle\varphi_{L}\rangle^2$. Then the first term in r.h.s. of Eq.
(\ref{16}) is reduced to a familiar Vlasov term,  and the second
term in r.h.s. of  Eq. (\ref{16})   is associated with correlators
of fluctuations induced by interactions with short-length modes. In
such an approximation, the above equation cannot describe
thermalization, but it still can describe process of momentum
isotropization  and spatiotemporal decoherence of the long-length
modes, which  precedes thermal equilibration acting on a shorter
time scale and is  a necessary condition for thermalization and
hydrodynamics.

Isotropization  of the relevant subsystem can happen, evidently,
only because of interactions with irrelevant modes. In more
mathematical terms, it can happen if the diffusion term, which  is
associated with correlators of fluctuations, has appropriate
properties. Namely, let us assume that $\tau^{\mu \nu}\sim
\delta^{\mu \nu}$, i.e., the corresponding fluctuations are
isotropic. Then, to find a steady (quasistationary) state, we
suppose that r.h.s. of Eq. (\ref{16}) is equal to zero:
\begin{eqnarray}
\frac{1}{2}
\partial^{0}\varrho (x) \frac{\partial}{\partial p_{0}^2} N_{L} (x,p)+  \frac{1}{4}
\frac{\partial}{\partial p_{0}^2} \frac{\partial}{\partial
p_{0}^2}\tau^{0 0} N_{L}(x,p) + \nonumber \\
\frac{1}{4p_{0}}
\partial^{i}\varrho (x) \frac{\partial}{\partial p^{i}} N_{L} (x,p) +
\frac{1}{16 p_{0}^2} \frac{\partial}{\partial p^{i}}
\frac{\partial}{\partial p^{i}}\tau^{ii} N_{L}(x,p) = 0. \label{17}
\end{eqnarray}
Here for convenience we divided  the r.h.s. of Eq. (\ref{16}) on
$p_0$. Just to demonstrate that the  solution of the above equation
can be associated with the isotropic steady state, let us find an
approximate analytic solution of Eq. (\ref{16}) for
$|\textbf{p}|/|p_{0}|\ll 1$. One can easily  see that it is
\begin{eqnarray}
 N_{L}(x,p)\sim \exp \left  [-2p_{0} \left ( \frac{p_{0}\partial^{0}\varrho (x)}{\tau^{00}}+
 \frac{2 p_{i}\partial^{i}\varrho (x)}{\tau^{ii}} \right ) \right ]. \label{18}
\end{eqnarray}
Notice, that  such a steady state is obtained without account of
energy-momentum dispersion relation. So, it is valid, in fact, only
if mass shell constraint on $p_{\mu}$ is not strongly peaked like
the  delta-function but, instead, is wide enough, having, however,
some limited virtuality.

For an expanding system one can expect that $\partial^{\mu}\varrho
>0$ because $\varrho <0$, see (\ref{10.0}) and (\ref{10.2}). Then,
this steady state has quasi - local equilibrium form with "effective
temperature" that is  $\sim 1/p_{0}$, and "effective collective
four-velocity" that is  $\sim
\partial^{\mu}\varrho$, and can be related to the so-called
prethermalization stage \cite{Berges}. Also, one can see that
(\ref{18}) demonstrates spatiotemporal decoherence  of the
long-wavelength subsystem. Indeed, the lengths of coherence are
associated with values of the off-diagonal elements of the
corresponding density matrix, $\rho_{L} (x+\frac{1}{2}\Delta
x,x-\frac{1}{2}\Delta x)=\int d^{4}p \exp (ip \Delta x)N_{L}(x,p)$.
The structure like (\ref{18}) for $N_{L}(x,p)$ leads typically to
finite coherence lengths. Because we do not fix the dispersion
relation, we just illustrate our conclusion analytically supposing
particles on a zero-mass shell. Then, calculating the density matrix
in the "rest frame",  it is easy to see that in the time direction
nondiagonal elements of the density matrix will be proportional to
$e^{-\Delta t^{2}/\lambda^{2}_{t}}$ with temporal correlation length
in this rest frame system to be $\lambda_{t} \propto
\sqrt{\partial^0\rho/\tau^{00}}$. During this time the long-length
state loses the coherence.  In space directions there are also
exponential cuts in nondiagonal elements of the density matrix. So
one can conclude that coherence lengths in this steady state are
finite and are caused by the   hydrodynamic-like parameters and
energy-momentum dispersion relation. In a similar way, thermal wave
length $\lambda_{th} \propto 1/\sqrt{mT}$  defines the off-diagonal
elements of the corresponding density matrix and, so, the spatial
coherence lengths of the non relativistic Boltzmann distribution.

One can see that (\ref{18}) is the isotropic expression in the
locally co-moving fluid-like rest frame. It is reasonable to expect
that for such a steady state the energy momentum tensor (\ref{9})
develops a sufficient degree of isotropy in the locally co-moving
frame. Because approximate isotropy in the locally co-moving frame
and decoherence of densities are  the basic premises for
applicability of hydrodynamics,  one can conjuncture that the energy
momentum tensor of the long-wavelength modes can, eventually,
approach to the energy momentum tensor of an effective viscous
fluid.

\section{Conclusions}

In this paper, we  discuss the physical mechanism that can explain
the source of decoherence at the early stage of matter evolution in
relativistic nucleus-nucleus collisions, and the subsequent approach
to hydrodynamical behavior. Our method, while admittedly heuristic,
provides a physical understanding of the decoherence phenomenon,
which was lacking in the current attempts of description of
thermalization in $A+A$ collisions, and sheds some light on the
mechanism of isotropization. In our opinion, understanding of the
dynamical mechanisms  of  decoherence and thermalization  should
create the necessary prerequisites  for unambiguous calculation of
viscous coefficients in $A+A$ collisions, and we hope that our
analysis can be useful for this aim.

Let us sum up our main points. First, it is well known that
decoherence of local densities is a necessary condition for
thermalization, and in an isolated system that is governed as whole
by the unitary quantum dynamics, the only possibility for
decoherence is decomposition of the system into subsystems.  An
ambiguity of a splitting procedure is removed by the requirement
that such a splitting must be done in an observer dependent way.
Taking into account typical observational conditions, we proposed to
split the system created in a relativistic $A+A$ collision into a
long-length modes subsystem and a short-length modes subsystem, and
consider the former as a relevant subsystem. Because the long-length
modes in the initial stage of a relativistic  $A+A$ collision are
highly populated \cite{CGC}, this  allows us to consider evolution
of the corresponding expectation value of the quantum field in the
quasiclassical approximation with noise term. The latter is
associated with quantum fluctuations that are mostly contributed by
the irrelevant (from an observational viewpoint) short-length
modes.\footnote{See also Ref. \cite{Giraud} where it was proposed
that unobservation of higher-order correlators may result in
effective decoherence and associated entropy production in quantum
field theory.} We suppose that such a stochasticity accounts
effectively for quantum entanglement between different scales. Then,
entanglement-driven stochasticity results in irreversibility and
decoherence for the effective coarse-grained dynamics of the large
scales.

We demonstrate how such a physical mechanism  works by means of
scalar field model.\footnote{ One can adjust this analysis for QCD
systems utilizing quark-gluon Wigner functions, see Ref.
\cite{Elze}.} We derived an evolution equation of the  Fokker-Planck
type for the Wigner function of the relevant part of the  system and
demonstrated, after some simplifying assumptions, that this equation
can describe decoherence and isotropization at prethermalization
stage, which  are necessary conditions for eventual thermalization
and hydrodynamics.   Notice that this  happens  as result of
interactions with the irrelevant (i.e., observationally
inaccessible) degrees of freedom, and  no averaging over the
ensemble of initial conditions is needed for such a quantum
thermalization. The generated non-zero entropy can be understood as
the entanglement entropy of the long-length subsystem of the system
created by a nucleus-nucleus collision, while the entropy of the
whole closed system does not change with time due to the unitarity
of the time evolution.

Our analysis supports the  idea that thermalization and transition
to hydrodynamics are contextual, and are related to a particular
procedure of decomposition of the whole quantum system into
subsystems that contain a large enough number degrees of freedom
(evidently, one should not expect a similar behavior in systems with
few degrees of freedom). Because observables are measured with some
degree of precision (and not all possible observables are measured)
in typical experiment, this  leaves the room for inaccessible
degrees of freedom, and, so, allows for hydrodynamical
approximation. A fluid dynamics then appears as an effective
long-wavelength theory. One can expect that the utilization of a
full unitary quantum evolution of a closed system with subsequent
projection into relevant coarse-grained subspace at the measurement
will result in the same predictions for a statistical ensemble of
experimental data as utilization  of a relevant coarse-grained
effective theory that follows to instantaneous decomposition of the
whole  state  into relevant and irrelevant subsystems/observables
(for more discussions see Ref. \cite{Paz-1}). On the other hand, a
hydrodynamical description is inappropriate for an observer who
wholly measures the total set of observables for an isolated quantum
system. Such an observer then will have to calculate the whole
quantum evolution of a system of  interest to predict results of
such a "complete" experiment.

\section{Acknowledgments}
Yu.S. thanks  J. Maldacena, A. Ashtekar, P. Braun-Munzinger for
useful discussions of the paper subject, and  ExtreMe Matter
Institute EMMI for the visiting professor position. The research was
carried out within the scope of the EUREA: European Ultra
Relativistic Energies Agreement (European Research Group: "Heavy
Ions at Ultrarelativistic Energies") and is supported by the
National Academy of Sciences of Ukraine (Agreement - 2014).

\end{document}